Pentaquarks in a relativistic quark model and nature of Theta-states.

Gerasyuta S.M.[1,2], Kochkin V.I.[1]

1. Department of Theoretical Physics, St. Petersburg State University, 198904, St. Petersburg, Russia.
2. Department of Physics, LTA, 194021, St. Petersburg, Russia,
   E-mail: gerasyuta@sg6488.spb.edu

Abstract

The relativistic five-quark equations are found in the framework of the dispersion relation technique. The solutions of these equations using the method based on the extraction of the leading singularities of the amplitudes are obtained. The five-quark amplitudes for the low-lying pentaquarks including the u, d, s- quarks are calculated. The poles of these amplitudes determine the masses of Theta-pentaquarks. The mass spectra of the isotensor Theta-pentaquarks with $J^P = \frac{1}{2}^\pm, \frac{3}{2}^\pm$ are calculated.

I. Introduction

Recently a number of laboratories have announced observation of a strangeness +1 baryon [1, 2, 3, 4, 5] with a mass of 1540 MeV and a narrow decay width. Such state cannot be as 3-quark baryon made from known quarks, and it is natural to interpret it as a pentaquark state. This state is made from four quarks and one antiquark $q^4\bar{q}$.

Experimental evidence was first reported by LEPS group at Spring-8 [1] in a reaction $\gamma n \to K^+ K^- n$. The current example of the strangeness S=+1 baryon is positively charged and called $\theta^+$. The $\theta^+$ of necessity has $\bar{s}$ and four nonstrange quarks. The parity, spin and isospin of the experimental state have not been measured yet.

Theoretically, using the chiral quark model, Diakonov et al. [6] have already studied the properties of $\theta^+$ as a member of antidecuplet with prediction of mass and width which are suprisingly close to the result experimentalists of observed. Futhermore, the spin and parity $J^P$ are predicted to be $J^P = \frac{1}{2}^+$ (isospin $I = 0$).

The recent announcements by the experimental groups have spawned a flurry of theoretical works [7 - 16] using such tools as quark model with bags, potentials, and pure group theory and chiral soliton models. The existence of light, exotic, positive parity baryons belonging to antidecuplet is a natural feature of chiral models [17, 19]. Hosaka emphasized the important role of the pion in the strong interaction dynamics which drives the formation of the $\theta^+$ particle and the detailed energy level ordering [8]. Capstick, Page and Roberts hypothesized that the $\theta^+$ is interpreted as a state containing a dominant pentaquark Fock-state component $uudd\bar{s}$ decaying to $K^+ n$, in contrast to its interpretation as a chiral soliton. In this pure multiquark picture $\theta^+$ has isospin, spin and parity different from those predicted by the Skyrme model [6], e.g. $\theta^+$ can be an isotensor resonance with $J^P = \frac{1}{2}^-, \frac{3}{2}^-, \frac{5}{2}^-$ [7].

Jaffe and Wilczek suggested that the observed $\theta^+$ state could be composed of an antistrange quark and two highly correlated up and down quark pairs arising from strong color-spin correlation force [9]. The resulting $J^P$ of $\theta^+$ state is $\frac{1}{2}^+$.

Karliner and Lipkin discussed the dynamics of diquark-triquark state with a rough estimate of the mass of 1592 MeV and isospin $I = 0$, $J^P = \frac{1}{2}^+$ in the constituent quark model [10]. Shi-Lin Zhu estimated the mass of the pentaquark state with QCD sum rules and found that the pentaquark state with isospin $I = 0$ lie close to 1550 MeV [11].

In our previous paper [20] the relativistic generalization of five-quark equations (like Faddeev-Yakubovsky approach) are constructed in the form of the dispersion relation. The five-quark amplitudes for the low-lying pentaquarks contain only u, d- quarks. The poles of these amplitudes determine the masses of the $ududd$ and $uuuu\bar{u}$ pentaquarks. The constituent quark is the color triplet and the quark amplitudes obey the global color symmetry. The interesting result of this model is the calculation of pentaquark amplitude which contain the contribution of four subamplitudes: molecular subamplitude $BM$, $D\bar{q}D$ subamplitude, $Mqqq$ subamplitude and $Dqq\bar{q}$ subamplitude. Here $B$ corresponds to the lowest baryon (nucleon and $\Delta$ - isobar baryon). $M$ are the low-lying mesons with the quantum numbers: $J^{PC} = 0^{++}, 1^{++}, 2^{++}, 0^{-+}, 1^{--}$ and $I_z = 0,1$. $D$ are the diquarks with $J^P = 0^+, 1^+$. The lowest mass of nucleon pentaquark with $J^P = \frac{1}{2}^+$ is equal M=1686 MeV.

The present paper is devoted to the construction of relativistic five-quark equations for the family of the $\theta$ pentaquarks. The five-quark amplitudes for the low-lying $\theta$ pentaquarks are calculated. The poles of these amplitudes determine the masses of the $\theta$ pentaquarks. The masses of the constituent u, d, s -quarks coincide with the quark masses of the ordinary baryons in our quark model [21]: $m_{u,d} = 410$ MeV, $m_s = 557$ MeV. We received the isotensor $\theta$ pentaquarks with $J^P = \frac{1}{2}^{\pm}, \frac{3}{2}^{\pm}$ and predict the masses of $\theta^{++}(\theta^0)$ and $\theta^{+++}(\theta^-)$ pentaquarks (Table 1). The interesting result of this model is the calculation of pentaquark amplitudes which contain the contribution of the subamplitudes for the $\theta^{+++}$, $\theta^{++}$ and $\theta^+$ states with negative parity:

$\theta^{+++}$ (Fig.1)   $K^+\Delta^{++}, K^+uuu$,

$\theta^{++}$ (Fig.2)   $K^+p, K^0\Delta^{++}, K^+uud$,

$\theta^+$ (Fig.3)   $K^0p, K^+n, K^+udd$.

The main contribution to the pentaquark amplitude is determined by the subamplitudes which include the mesons with $J^{PC} = 0^{++}, 0^{-+}$.

The paper is organized as follows. After this introduction, we discuss the five-quark amplitudes which contain $\bar{s}$ - antiquark and four nonstrange quarks (Section 2). In section 3, we report our numerical results (Table 1) and the last section is devoted to the discussion and conclusion.

## II. Five-quark amplitudes for the family of exotic $\theta$ - baryons

We derive the relativistic five-quark equations in the framework of the dispersion relation technique. We use only planar diagrams; the other diagrams due to the rules of $1/N_c$ expansion [22 - 24] are neglected. The correct equations for the amplitude are obtained by taking into account all possible subamplitudes. It corresponds to the division of complete system into subsystems smaller number of particles. Then one should represent a five-particle amplitude as a sum of ten subamplitudes: $A = A_{12} + A_{13} + A_{14} + A_{15} + A_{23} + A_{24} + A_{25} + A_{34} + A_{35} + A_{45}$.

We need to consider only one group of diagrams and the amplitude corresponding to them, for example $A_{12}$. For the sake of simplicity we shall consider the derivation of the relativistic generalization of the Faddeev-Yakubovsky approach for the example of $\theta^{+++}$ pentaquark. We shall construct the five-quark amplitude of four up quarks and one strange antiquark, in which only pair interaction with the quantum numbers of a $J^P = 1^+$ diquark are included. The set of diagrams associated with the amplitude $A_{12}$ can further be broken down into groups correspondin to amplitudes: $A_1(s, s_{1234}, s_{12}, s_{34})$, $A_2(s, s_{1234}, s_{25}, s_{34})$, $A_3(s, s_{1234}, s_{13}, s_{134})$, $A_4(s, s_{1234}, s_{24}, s_{234})$ (Fig. 1). The $\bar{s}$ is shown by the arrow and other lines correspond to the four nonstrange quarks. In the case of the $\theta^{++}(\theta^0)$ and $\theta^+$ pentaquarks we need to use six (Fig.2) or seven (Fig.3) subamplitudes. The coefficients are determined by the permutation of quarks [25, 26].

In order to represent the subamplitudes $A_1(s, s_{1234}, s_{12}, s_{34})$, $A_2(s, s_{1234}, s_{25}, s_{34})$, $A_3(s, s_{1234}, s_{13}, s_{134})$ and $A_4(s, s_{1234}, s_{24}, s_{234})$ in the form of a dispersion relation it is necessary to define the amplitudes of quark-quark and quark-antiquark interaction $b_n(s_{ik})$. The pair quarks amplitudes $q\bar{q} \to q\bar{q}$ and $qq \to qq$ are calculated in the framework of the dispersion

N/D method with the input four-fermion interaction with quantum numbers of the gluon [27]. We use the results of our relativistic quark model [27] and write down the pair quarks amplitude in the form:

$$b_n(s_{ik}) = \frac{G_n^2(s_{ik})}{1 - B_n(s_{ik})}, \quad (1)$$

$$B_n(s_{ik}) = \int_{(m_1+m_2)^2}^{\Lambda_n} \frac{ds'_{ik}}{\pi} \frac{\rho_n(s'_{ik}) G_n^2(s'_{ik})}{s'_{ik} - s_{ik}}. \quad (2)$$

Here $s_{ik}$ is the two-particle subenergy squared, $s_{ijk}$ corresponds to the energy squared of particles $i$, $j$, $k$, $s_{ijkl}$ is the four-particle subenergy squared and $s$ is the system total energy squared. $G_n(s_{ik})$ are the quark-quark and quark-antiquark vertex functions (Table 2). $B_n(s_{ik})$, $\rho_n(s_{ik})$ are the Chew-Mandelstam function with cut-off $\Lambda_n$ ($\Lambda_1 = \Lambda_3$) [20] and the phase space respectively:

$$\rho_n(s_{ik}, J^{PC}) = \left( \alpha(J^{PC}, n) \frac{s_{ik}}{(m_i + m_k)^2} + \beta(J^{PC}, n) \right) \frac{\sqrt{[s_{ik} - (m_i + m_k)^2][s_{ik} - (m_i - m_k)^2]}}{s_{ik}},$$

The coefficients $\alpha(J^{PC}, n)$ and $\beta(J^{PC}, n)$ are given in Table 3. Here n=1 corresponds to a $qq$-pair with $J^P = 0^+$ in the $\bar{3}_c$ color state, n=2 describes a $qq$-pair with $J^P = 1^+$ in the $\bar{3}_c$ color state and n=3 defines the $q\bar{q}$-pairs corresponding to mesons with quantum numbers: $J^{PC} = 0^{++}, 0^{-+}$.

In the case in question the interacting quarks do not produce a bound state, therefore the integration in Eqs. (3) - (6) is carried out from the threshold $(m_i + m_k)^2$ to the cut-off $\Lambda_n$. The system of integral equations systems, corresponding to Fig. 1 (the meson state with $J^{PC} = 0^{++}$ and diquark with $J^P = 1^+$) can be described as:

$$A_1(s, s_{1234}, s_{12}, s_{34}) = \frac{\lambda_1 B_3(s_{12}) B_2(s_{34})}{[1 - B_3(s_{12})][1 - B_2(s_{34})]} + 6\hat{J}_2(3,2) A_4(s, s_{1234}, s'_{23}, s'_{234}) +$$
$$+ 2\hat{J}_2(3,2) A_3(s, s_{1234}, s'_{13}, s'_{134}) + 2\hat{J}_1(3) A_3(s, s_{1234}, s'_{15}, s_{125}) + 2\hat{J}_1(2) A_4(s, s_{1234}, s'_{25}, s_{125}) +, \quad (3)$$
$$+ 4\hat{J}_1(2) A_4(s, s_{1234}, s'_{35}, s_{345})$$

$$A_2(s, s_{1234}, s_{25}, s_{34}) = \frac{\lambda_2 B_2(s_{25}) B_2(s_{34})}{[1 - B_2(s_{25})][1 - B_2(s_{34})]} +$$
$$+ 12 \hat{J}_2(2,2) A_4(s, s_{1234}, s'_{23}, s'_{234}) + 8 \hat{J}_1(2) A_3(s, s_{1234}, s'_{25}, s_{125}) \quad (4)$$

$$A_3(s, s_{1234}, s_{13}, s_{134}) = \frac{\lambda_3 B_3(s_{12})}{1 - B_3(s_{12})} + 12 \hat{J}_3(3) A_1(s, s_{1234}, s'_{12}, s'_{34}), \quad (5)$$

$$A_4(s, s_{1234}, s_{24}, s_{234}) = \frac{\lambda_4 B_2(s_{24})}{1 - B_2(s_{24})} + 4 \hat{J}_3(2) A_2(s, s_{1234}, s'_{25}, s'_{34}) + 4 \hat{J}_3(2) A_1(s, s_{1234}, s'_{12}, s'_{34}), \quad (6)$$

were $\lambda_i$ are the current constants. We introduce the integral operators:

$$\hat{J}_1(l) = \frac{G_l(s_{12})}{[1 - B_l(s_{12})]} \int_{(m_1+m_2)^2}^{\Lambda_l} \frac{ds'_{12}}{\pi} \frac{G_l(s'_{12}) \rho_l(s'_{12})}{s'_{12} - s_{12}} \int_{-1}^{+1} \frac{dz_1}{2}, \quad (7)$$

$$\hat{J}_2(l,p) = \frac{G_l(s_{12}) G_p(s_{34})}{[1 - B_l(s_{12})][1 - B_p(s_{34})]} \times$$
$$\times \int_{(m_1+m_2)^2}^{\Lambda_l} \frac{ds'_{12}}{\pi} \frac{G_l(s'_{12}) \rho_l(s'_{12})}{s'_{12} - s_{12}} \int_{(m_3+m_4)^2}^{\Lambda_p} \frac{ds'_{34}}{\pi} \frac{G_p(s'_{34}) \rho_p(s'_{34})}{s'_{34} - s_{34}} \int_{-1}^{+1} \frac{dz_3}{2} \int_{-1}^{+1} \frac{dz_4}{2}, \quad (8)$$

$$\hat{J}_3(l) = \frac{G_l(s_{12}, \tilde{\Lambda})}{1 - B_l(s_{12}, \tilde{\Lambda})} \times$$
$$\times \frac{1}{4\pi} \int_{(m_1+m_2)^2}^{\tilde{\Lambda}} \frac{ds'_{12}}{\pi} \frac{G_l(s'_{12}, \tilde{\Lambda}) \rho_l(s'_{12})}{s'_{12} - s_{12}} \int_{-1}^{+1} \frac{dz_1}{2} \int_{-1}^{+1} dz \int_{z_2^-}^{z_2^+} dz_2 \frac{1}{\sqrt{1 - z^2 - z_1^2 - z_2^2 + 2 z z_1 z_2}}, \quad (9)$$

were $l, p$ are equal 1 - 3. Here $m_i$ is a quark mass.

In the Eqs. (7) and (9) $z_1$ is the cosine of the angle between the relative momentum of the particles 1 and 2 in the intermediate state and the momentum of the particle 3 in the final state, taken in the c.m. of particles 1 and 2. In the Eq. (9) $z$ is the cosine of the angle between the momenta of the particles 3 and 4 in the final state, taken in the c.m. of particles 1 and 2. $z_2$ is the cosine of the angle between the relative momentum of particles 1 and 2 in the intermediate state and the momentum of the particle 4 in the final state, is taken in the c.m. of particles 1 and 2. In the Eq. (8): $z_3$ is the cosine of the angle between relative momentum of particles 1 and 2 in the intermediate state and the relative momentum of particles 3 and 4 in the intermediate state, taken in the c.m. of particles 1 and 2. $z_4$ is the cosine of the angle between the relative momentum of the particles 3 and 4 in the intermediate state and that of the momentum of the particle 1 in the intermediate state, taken in the c.m. of particles 3, 4.

We can pass from the integration over the cosines of the angles to the integration over the subenergies [28].

Let us extract two-particle singularities in the amplitudes $A_1(s, s_{1234}, s_{12}, s_{34})$, $A_2(s, s_{1234}, s_{25}, s_{34})$, $A_3(s, s_{1234}, s_{13}, s_{134})$ and $A_4(s, s_{1234}, s_{24}, s_{234})$:

$$A_1(s, s_{1234}, s_{12}, s_{34}) = \frac{\alpha_1(s, s_{1234}, s_{12}, s_{34}) B_3(s_{12}) B_2(s_{34})}{[1 - B_3(s_{12})][1 - B_2(s_{34})]}, \tag{10}$$

$$A_2(s, s_{1234}, s_{25}, s_{34}) = \frac{\alpha_2(s, s_{1234}, s_{25}, s_{34}) B_2(s_{25}) B_2(s_{34})}{[1 - B_2(s_{25})][1 - B_2(s_{34})]}, \tag{11}$$

$$A_3(s, s_{1234}, s_{13}, s_{134}) = \frac{\alpha_3(s, s_{1234}, s_{13}, s_{134}) B_3(s_{13})}{1 - B_3(s_{13})}, \tag{12}$$

$$A_4(s, s_{1234}, s_{24}, s_{234}) = \frac{\alpha_4(s, s_{1234}, s_{24}, s_{234}) B_2(s_{24})}{1 - B_2(s_{24})}. \tag{13}$$

We do not extract three- and four-particle singularities, because they are weaker than two-particle singularities.

We used the classification of singularities, which was proposed in paper [20]. The construction of approximate solution of Eqs. (3) - (6) is based on the extraction of the leading singularities of the amplitudes. The main singularities in $s_{ik} \approx (m_i + m_k)^2$ are from pair rescattering of the particles i and k. First of all there are threshold square-root singularities. Also possible are pole singularities which correspond to the bound states. The diagrams of Fig.1 apart from two-particle singularities have the triangular singularities, the singularities defining the interaction of four and five particles. Such classification allows us to search the corresponding solution of Eqs. (3) - (6) by taking into account some definite number of leading singularities and neglecting all the weaker ones. We consider the approximation which defines two-particle, triangle, four- and five-particle singularities. The functions $\alpha_1(s, s_{1234}, s_{12}, s_{34})$, $\alpha_2(s, s_{1234}, s_{25}, s_{34})$, $\alpha_3(s, s_{1234}, s_{13}, s_{134})$ and $\alpha_4(s, s_{1234}, s_{24}, s_{234})$ are smooth functions of $s_{ik}$, $s_{ijk}$, $s_{ijkl}$, $s$ as compared with the singular part of the amplitudes, hence they can be expanded in a series in the singularity point and only the first term of this series should be employed further. Using this classification one define the reduced amplitudes $\alpha_1$, $\alpha_2$, $\alpha_3$, $\alpha_4$ as well as the B-functions in the middle point of the physical region of Dalitz-plot at the point $s_0$:

$$s_0^{ik} = s_0 = \frac{s + 3\sum_{i=1}^{5} m_i^2}{0.25 \sum_{\substack{i,k=1 \\ i \neq k}}^{5} (m_i + m_k)^2} \tag{14}$$

$$s_{123} = 0.25 s_0 \sum_{\substack{i,k=1 \\ i \neq k}}^{3} (m_i + m_k)^2 - \sum_{i=1}^{3} m_i^2, \quad s_{1234} = 0.25 s_0 \sum_{\substack{i,k=1 \\ i \neq k}}^{4} (m_i + m_k)^2 - 2\sum_{i=1}^{4} m_i^2$$

Such a choice of point $s_0$ allows one to replace the integral Eqs. (3) - (6) (Fig. 1) by the algebraic equations (16) - (19) respectively:

$$\alpha_1 = \lambda_1 + 6\alpha_4 J_2(3,2,2) + 2\alpha_3 J_2(3,2,3) + 2\alpha_3 J_1(3,3) + 2\alpha_4 J_1(3,2) + 4\alpha_4 J_1(2,2), \tag{15}$$
$$\alpha_2 = \lambda_2 + 12\alpha_4 J_2(2,2,2) + 8\alpha_3 J_1(2,3), \tag{16}$$
$$\alpha_3 = \lambda_3 + 12\alpha_1 J_3(3,3,2), \tag{17}$$
$$\alpha_4 = \lambda_4 + 4\alpha_2 J_3(2,2,2) + 4\alpha_1 J_3(2,2,3). \tag{18}$$

We use the functions $J_1(l,p)$, $J_2(l,p,r)$, $J_3(l,p,r)$ ($l,p,r = 1, 2, 3$):

$$J_1(l,p) = \frac{G_l^2(s_0^{12}) B_p(s_0^{13})}{B_l(s_0^{12})} \int_{(m_1+m_2)^2}^{\Lambda_l} \frac{ds'_{12}}{\pi} \frac{\rho_l(s'_{12})}{s'_{12} - s_0^{12}} \int_{-1}^{+1} \frac{dz_1}{2} \frac{1}{1 - B_p(s'_{13})}, \tag{19}$$

$$J_2(l,p,r) = \frac{G_l^2(s_0^{12}) G_p^2(s_0^{34}) B_r(s_0^{13})}{B_l(s_0^{12}) B_p(s_0^{34})} \times$$
$$\times \int_{(m_1+m_2)^2}^{\Lambda_l} \frac{ds'_{12}}{\pi} \frac{\rho_l(s'_{12})}{s'_{12} - s_0^{12}} \int_{(m_3+m_4)^2}^{\Lambda_p} \frac{ds'_{34}}{\pi} \frac{\rho_p(s'_{34})}{s'_{34} - s_0^{34}} \int_{-1}^{+1} \frac{dz_3}{2} \int_{-1}^{+1} \frac{dz_4}{2} \frac{1}{1 - B_r(s'_{13})} \tag{20}$$

$$J_3(l,p,r) = \frac{G_l^2(s_0^{12}, \tilde{\Lambda}) B_p(s_0^{13}) B_r(s_0^{24})}{1 - B_l(s_0^{12}, \tilde{\Lambda})} \frac{1 - B_l(s_0^{12})}{B_l(s_0^{12})} \times$$
$$\times \frac{1}{4\pi} \int_{(m_1+m_2)^2}^{\tilde{\Lambda}} \frac{ds'_{12}}{\pi} \frac{\rho_l(s'_{12})}{s'_{12} - s_0^{12}} \int_{-1}^{+1} \frac{dz_1}{2} \int_{-1}^{+1} dz \int_{z_2^-}^{z_2^+} dz_2 \frac{1}{\sqrt{1 - z^2 - z_1^2 - z_2^2 + 2zz_1z_2}} \frac{1}{[1 - B_p(s'_{13})][1 - B_r(s'_{24})]} \tag{21}$$

The other choices of point $s_0$ do not change essentially the contributions of $\alpha_1$, $\alpha_2$, $\alpha_3$ and $\alpha_4$, therefore we omit the indexes $s_0^{ik}$. Since the vertex functions depend only slightly on energy it is possible to treat them as constants in our approximation. The integration contours of function $J_1, J_2, J_3$ are given in paper [28].

The solutions of the system of equations are considered as:

$$\alpha_i(s) = F_i(s, \lambda_i)/D(s), \qquad (22)$$

where zeros of $D(s)$ determinants define the masses of bound states of pentaquark baryons. $F_i(s, \lambda_i)$ are the functions of $s$ and $\lambda_i$. The functions $F_i(s, \lambda_i)$ determine the contributions of subamplitudes to the pentaquark baryon amplitude.

### III. Calculation results

The poles of the reduced amplitudes $\alpha_1$, $\alpha_2$, $\alpha_3$, $\alpha_4$ correspond to the bound states and determine the masses of $\theta^{+++}$ ($\theta^-$) pentaquarks (Fig.1). If we consider the $\theta^{++}$ ($\theta^0$) (Fig.2) or $\theta^+$ (Fig.3) states, we must take into account the interaction of the quarks in the $0^+$ and $1^+$ states. The quark masses of model $m_{u,d} = 410$ MeV and $m_s = 557$ MeV coincide with the quark masses of the ordinary baryons in our model [21]. The model in consideration has only one new parameter as compared to previous paper [20]. The gluon coupling constant $g = 0.456$ is determined by fixing of $\theta^+$ pentaquark mass (1540 MeV). The cut-off parameters coincide with those in paper [20] : $\Lambda_{0^+} = 16.5$ and $\Lambda_{1^+} = 20.12$ for the diquarks with $0^+$ and $1^+$ respectively. The cut-off parameter for the mesons is equal to $\Lambda = 16.5$. The calculated mass values of low-lying isotensor pentaquarks $\theta$ are shown in Table1. We used the mass of lowest pentaquark $\theta^+$(1540) as fit. The masses of $\theta^{++}$ ($\theta^0$), $\theta^{+++}$ ($\theta^-$) pentaquarks with the quantum numbers $I_z = \pm 1, \pm 2$ and $J^P = \frac{1}{2}^\pm, \frac{3}{2}^\pm$ are predicted (Table1). The mass of $\theta$ pentaquarks with positive parity is smaller than the mass of pentaquarks with negative parity. It dependens on the different interaction in the diquark channels. We predict the degeneracy of $\theta^{+++}$ ($\theta^-$) states.

## IV. Conclusion

In strongly bound system of light quarks such as the baryons where $p/m \sim 1$, the approximation of nonrelativistic kinematics and dynamics is not justified. In our relativistic five-quark model (Faddeev-Yakubovsky type approach) the masses of family $\theta$ pentaquarks are calculated. We considered the scattering amplitudes of constituent quarks. The poles of these amplitudes determine the masses of low-lying pentaquarks: $\theta^+$, $\theta^{++}$ ($\theta^0$) and $\theta^{+++}$ ($\theta^-$). Capstick, Page and Roberts [7] proposed the most promising process for the production of the $\theta^{+++}$ state in reaction $pp \to \theta^{+++}\Sigma^-$, which involves one $s\bar{s}$ pair creation. Another promising process is $K^+ p \to \theta^{+++}\pi$ - involving the creation of one light quark pair. The search may be feasible at ITEP, JHF or COSY. If discovered, the $\theta^{+++}$ would be the first triply charged hadron. The states $\theta^{+++}$ ($\theta^-$) must decay weakly. The central members of isotensor ($\theta^0, \theta^+, \theta^{++}$) decay strongly. They have the isospin-violatily strong decays. We can consider the mass spectra of the pentaquarks with strangeness S=0,-1,-2,-3.

The interesting research is the consideration of the isotensor $qqqq\bar{Q}$ states with $q$ an up or down quark and $Q$ a heavy quark ($Q=c,b$). Their decay to $N(q\bar{Q})$ also violates isospin conservation. It is similar to the case of $\theta$ pentaquarks.

Acknowledgments

The authors would like to thank T. Barnes, D.I. Diakonov, A. Hosaka, N.N. Nikolaev, B. Silvestre-Brac for useful discussions. This research was supported in part by the Russian Ministry of Education, Program "Universities of Russia" under Contract N 01.20.00.06448.

Table I. Low-lying $\theta$ pentaquark masses (MeV)

| | $J^P$ | Mass, MeV |
|---|---|---|
| $\theta^{+++}$ ($uuuu\bar{s}$) $\theta^{-}$ ($dddd\bar{s}$) | $\frac{1}{2}^+, \frac{3}{2}^+$ | 1727 |
| | $\frac{5}{2}^+$ | - |
| | $\frac{1}{2}^-, \frac{3}{2}^-$ | 1704 |
| $\theta^{++}$ ($uuud\bar{s}$) $\theta^{0}$ ($dddu\bar{s}$) | $\frac{1}{2}^+$ | 1575 |
| | $\frac{3}{2}^+$ | 1761 |
| | $\frac{1}{2}^-$ | 1630 |
| | $\frac{3}{2}^-$ | 1857 |
| $\theta^{+}$ ($udud\bar{s}$) | $\frac{1}{2}^+$ | 1540 |
| | $\frac{3}{2}^+$ | 1969 |
| | $\frac{1}{2}^-$ | 1643 |
| | $\frac{3}{2}^-$ | - |

Parameters of model: quark mass $m_{u,d}$ = 410 MeV, $m_s$ = 557 MeV;
cut-off parameter $\Lambda_{0^+}$ =16.5, $\Lambda_{1^+}$ =20.12; gluon coupling constant $g$ =0.456.

Table II. Vertex functions

| $J^{PC}$ | $G_n^2$ |
|---|---|
| $0^+$ (n=1) | $4g/3 - 2g(m_i + m_k)^2/(3s_{ik})$ |
| $1^+$ (n=2) | $2g/3$ |
| $0^{++}$ (n=3) | $8g/3$ |
| $0^{-+}$ (n=3) | $8g/3 - 4g(m_i + m_k)^2/(3s_{ik})$ |

Table III. Coefficient of Chew-Mandelstam functions for n = 3 (meson states) and diquarks n = 1 ($J^P = 0^+$), n = 2 ($J^P = 1^+$).

| $J^{PC}$ | n | $\alpha(J^{PC}, n)$ | $\beta(J^{PC}, n)$ |
|---|---|---|---|
| $0^{++}$ | 3 | ½ | -1/2 |
| $0^{-+}$ | 3 | ½ | -e/2 |
| $0^+$ | 1 | ½ | -e/2 |
| $1^+$ | 2 | 1/3 | 1/6-e/3 |

$e = (m_i - m_k)^2 / (m_i + m_k)^2$

Figure captions

Fig.1. Graphic representation of the equations for the five-quark subamplitudes $A_k$ ($k$=1-4) in the case of $\theta^{+++}(\theta^-)$ pentaquarks.

Fig.2. Graphic representation of the equations for the five-quark subamplitudes $A_k$ ($k$=1-6) corresponding to the $\theta^{++}(\theta^0)$ pentaquarks.

Fig.3. Graphic representation of the equations for the five-quark subamplitudes $A_k$ ($k$=1-7) corresponding to the $\theta^+$ pentaquark.

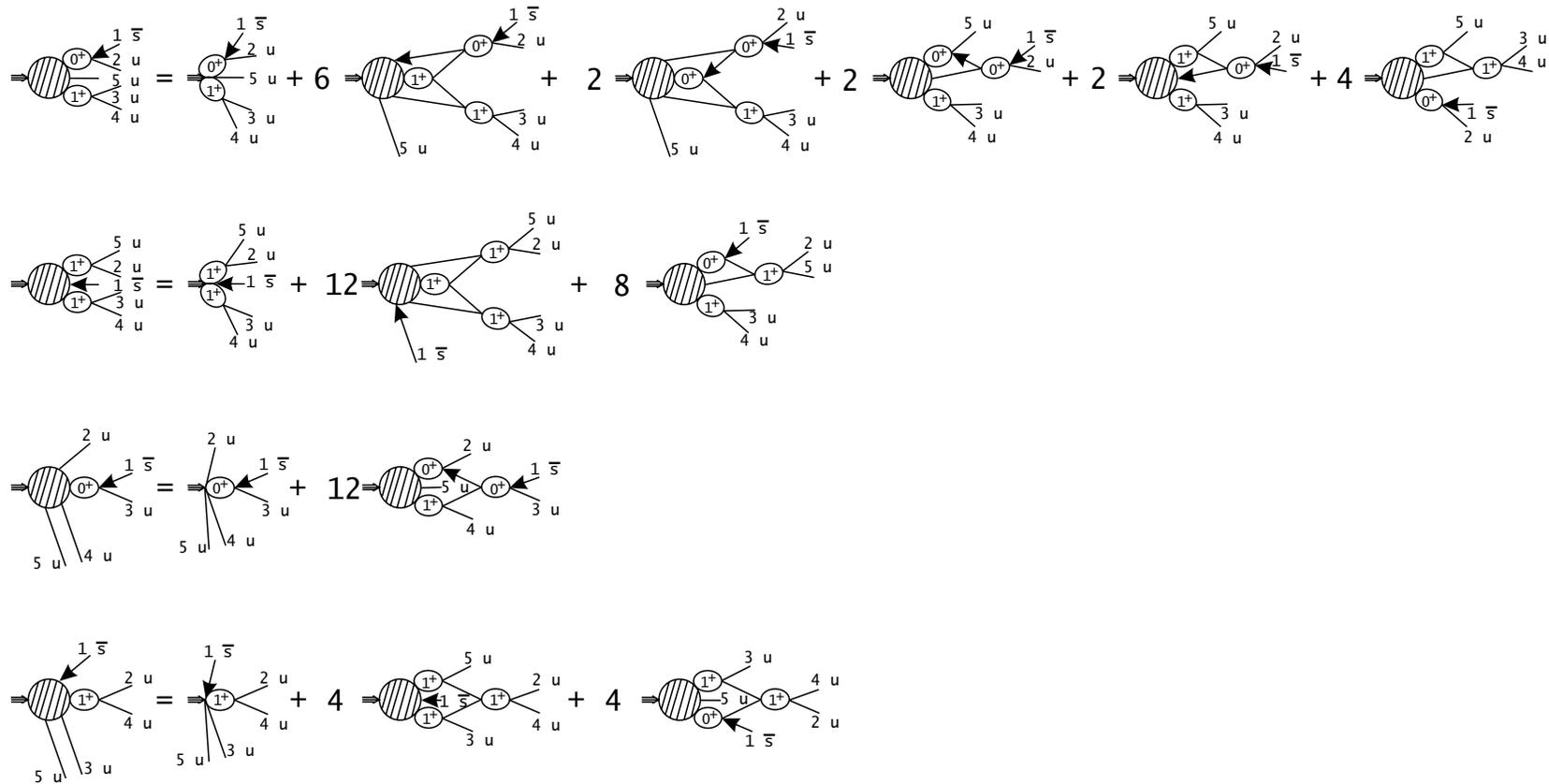

Fig.1

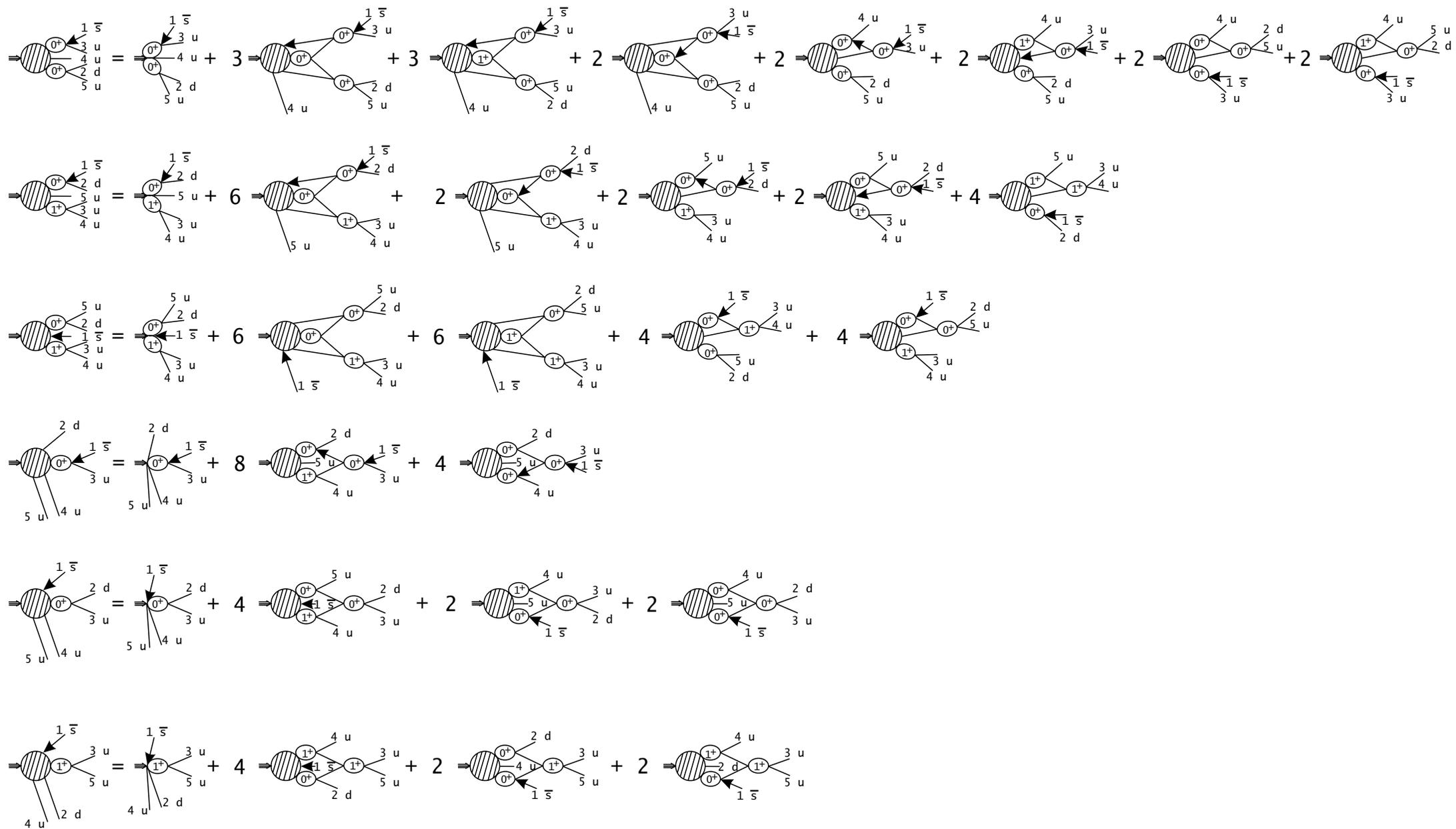

Fig.2

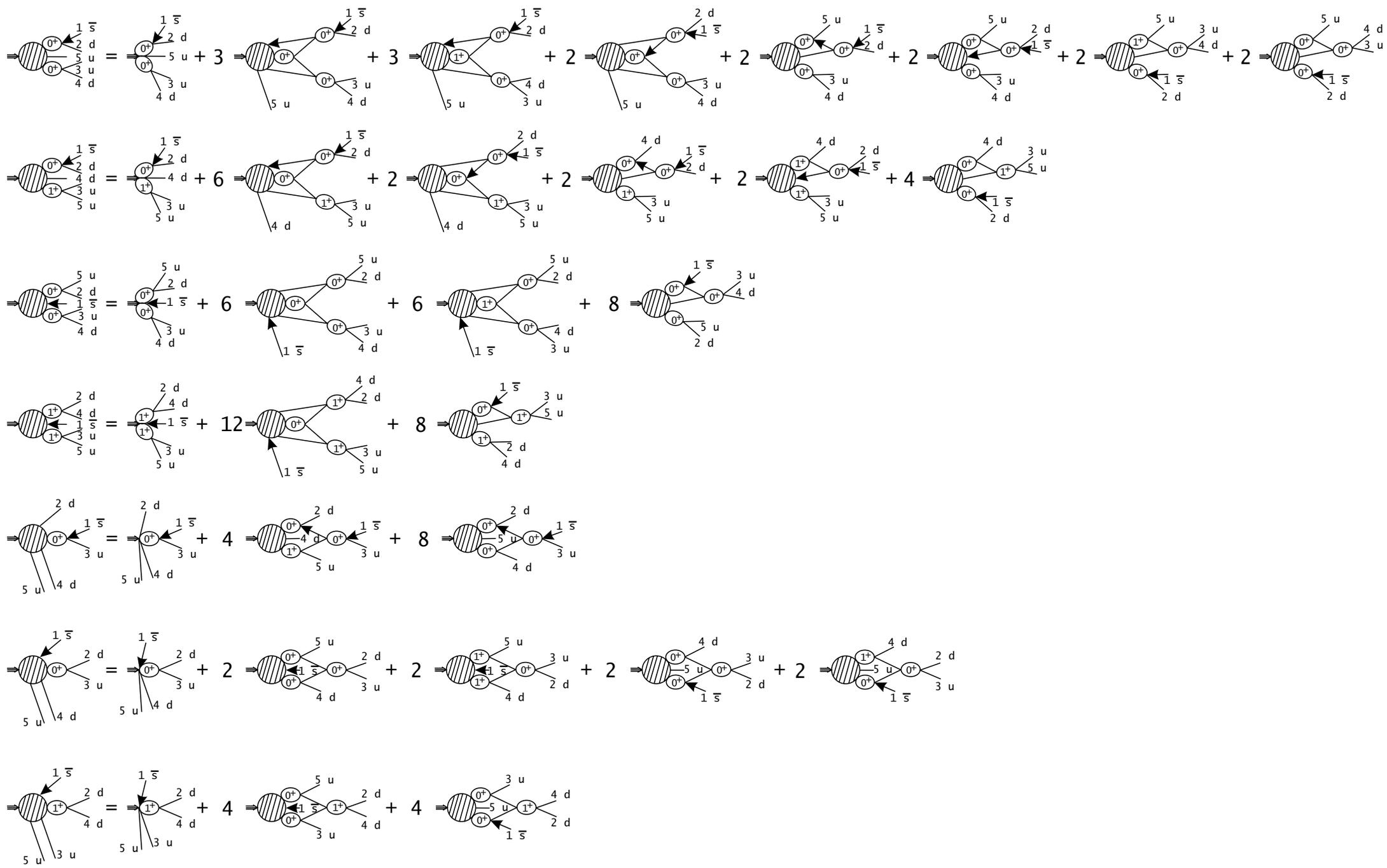

Fig.3